\documentclass[reprint,aps,prc,amsmath,amssymb,showpacs,showkeys,twocolumns,floatfix,diagbox]{revtex4-1}
\usepackage{graphicx}
\usepackage{dcolumn}
\usepackage{bm}
\usepackage{longtable}
\usepackage[flushleft]{threeparttable}
\usepackage{hyperref}
\usepackage{siunitx}
\usepackage{float}
\usepackage{tablefootnote}
\begin{document}

\preprint{APS/123-QED}

\title[$\alpha$ decay law of excited nuclei and its role in stellar decay rates]{$\alpha$ decay law of excited nuclei and its role in stellar decay rates}

\author{D. F. Rojas-Gamboa}
\email{df.rojas11@uniandes.edu.co}
\affiliation{Departamento de F\'isica, Universidad de Los Andes,
Carrera 1 No. 18 A - 10, Bogot\'a 111711, Colombia}
\author{N. G. Kelkar}
\email{nkelkar@uniandes.edu.co}
\affiliation{Departamento de F\'isica, Universidad de Los Andes,
Carrera 1 No. 18 A - 10, Bogot\'a 111711, Colombia}
\author{O. L. Caballero}
\email{ocaballe@uoguelph.ca}
\affiliation{Department of Physics, University of Guelph, Guelph, ON N1G 2W1, Canada}

\begin{abstract}
$\alpha$ decay is one of the prominent decay modes in the nucleosynthesis of heavy and super-heavy elements synthesized at temperatures of the order of Giga Kelvin. To facilitate the investigation of the role played by the $\alpha$ decay half-lives of thermally excited nuclei in nucleosynthesis calculations, an empirical formula based on a model for the $\alpha$ decay of nuclei in their ground and excited states to daughter nuclei in their ground or excited states is presented. Constants appearing in the analytical expression for the $\alpha$ decay half-life obtained within the model are treated as adjustable parameters and fitted to experimental data on 342 $\alpha$ decays in the range of  82 $\le Z_p \le$ 94, to obtain an excitation energy-dependent decay law. Under the assumption that thermal equilibrium has been reached between nuclear states, temperature ($T$)-dependent half-lives, $t_{1/2}(T)$, for several of the experimentally studied $\alpha$ emitters with 65 $\le Z_p \le$ 94 are presented using available data on the half-lives of excited nuclei. Though the general trend is a decrease in $t_{1/2}(T)$ at elevated temperatures, exceptional cases with increased half-lives are found in the case of some isomeric states. A list of such isomers provided in this work motivates future work involving considerations of their thermal equilibration and role in shaping kilonova light curves.   
\end{abstract}



\maketitle

\section{Introduction}\label{intro}
Empirical laws are not uncommon in physics, especially in nuclear physics where the many-body problem makes it difficult to include the properties of the nuclei and the interactions of the constituents in a microscopic formalism. Semi-empirical formulas based on simple models with parameters fitted to existing data indeed prove to be a useful tool in understanding nuclear phenomena. One of the oldest examples is the semi-empirical mass formula \cite{Bethe1936,Weizsacker1935,Seeger1961} based on the liquid drop model of the nucleus. This simple formula that estimated nuclei's binding energies or masses based on the number of protons and neutrons was refined and reused over the years in literature. Improved versions such as the finite range droplet model \cite{Moller1981} combined with microscopic contributions based on a folded Yukawa single particle potential \cite{Moller1988,Moller2016} proved useful in providing more precise nuclear masses as well as ground state deformations of nuclei. For applications in astrophysics and heavy ion collisions, the simple formula was extended to non-zero excitation energies \cite{Davidson1994}. Another well-known example of an empirical formula is the Geiger-Nuttall (GN) law \cite{GeigerNuttall1911}, which relates the half-lives in $\alpha$ decay of nuclei to the $Q$-values or the energies of the $\alpha$ particles emitted in the decay of radioactive nuclei. A century later, the Geiger-Nuttall law,
\begin{equation}\label{GeigerN}
    \log_{10} t_{1/2} = a(Z) Q^{-1/2} + b(Z) \,, 
\end{equation} 
continues to attract the attention of the community \cite{QiPLB2014}. With increasing data that has become available over the hundred years, the original GN law is not sufficient to reproduce all $\alpha$ decay half-lives. Hence, modifications of this law with additional inputs are proposed \cite{Ren2012}. In 1966, Viola and Seaborg (VS), generalized the GN formula and included an additional term depending on the so-called hindrance factor \cite{Viola1966}. New fits to this formula were performed in Ref. \cite{Royer2000,Dong2005}. Since these empirical decay laws can be explained based on the tunneling effect in quantum mechanics, they have gained popularity and several ``universal decay laws'' which are extensions of the original GN and VS formulae are applied not only to $\alpha$ decay but also to cluster radioactivity and emission of charged particles in general \cite{Qi-etal2009, Delion2009,Soylu2021}.

In the present work, we intend to study the $\alpha$ decay of nuclei in their ground and excited states, decaying to daughter nuclei which can also either be in a ground or excited state, using a semi-empirical formula. In addition to the usual quantities such as the $Q$-value and the number of nucleons appearing in the fitted formula mentioned in the literature above, such a formula should depend on the excitation energies of the nuclei involved. Except for \cite{Delion2015}, where a generalized VS formula for half-lives that depended on the excitation energy of the daughter nucleus was given, most empirical formulae for half-lives predict only the ground state $\alpha$ decay half-lives of nuclei. In an earlier work \cite{RojasKelkar2022}, a decay law for the cluster radioactivity of excited nuclei \cite{DiegoThesis} was formulated by the present authors. The formula was motivated through an analytically solvable model for tunneling a charged light nucleus through the Coulomb barrier formed due to its interaction with the heavy daughter nucleus. Here, we perform some modifications to this formula and fit the parameters (which have otherwise a physical meaning in the model) to available data on $\alpha$ decay, for nuclei with (parent) atomic number $82\le Z_p\le 94$, where either the parent or daughter or both the parent and the daughter can be in an excited state. The semi-empirical formula obtained in this work can be useful for the calculation of the $\alpha$ decay half-lives at elevated temperatures in astrophysical environments \cite{Perrone1971, Jhoan_2023,Mohr_2023,Jhoan_2023_2}, where elements in that mass range are produced.

This article is organized as follows: in section \ref{formalism} we describe the steps we took to build our proposed model for $\alpha$ decay from and to excited states. In section \ref{Fittingdata} we adjust this model to a decay law with parameters fitted to evaluated experimental data and present the resulting coefficients and fit quality. The correlation matrix of the variables of the model is presented in the next subsection. We reformulate previous works on $\alpha$ decay from ground state to ground state, by adding an excitation energy dependence, and we compare with our results in subsections \ref{decay-laws}, and \ref{comparison-laws}. In section \ref{temperature}, we study the behavior of half-lives, $t_{1/2}^{\rm{\,exp}}(T)$ with temperature using the available experimental data on the $\alpha$ decay half-lives of excited nuclei. We note several interesting features of the temperature dependence of $\alpha$ decay. The role of isomeric states and thermal equilibration is also discussed in this section. Later in this section, the excitation energy-dependent decay law developed in the earlier sections is used to evaluate the $\alpha$ decay half-lives of excited nuclei which are in turn used to predict the temperature-dependent half-lives, $t_{1/2}^{\rm{\,cal}}(T)$. The model is validated by presenting the mean square error (MSE) of the difference between $t_{1/2}^{\rm{\,cal}}(T)$ and $t_{1/2}^{\rm{\,exp}}(T)$. In section \ref{scenarios}, astrophysical sites and conditions for the occurrence of $\alpha$ decay are presented.  In section \ref{conclusion} we give a summary and outlook.

\section{Model for $\alpha$ decay of excited nuclei} \label{formalism}
The relationship between the decay constant and quantities such as penetration probability, \textbf{$P$}, assault frequency, $\nu$, and $\alpha$ preformation factor, $P_{\alpha}$, allows us to write a logarithmic formula for the half-life, which proves convenient for deriving a parameterized analytical expression. We begin this process by formulating the following equation:
\begin{equation}\label{eq:UDL_T0}
    \log_{10}{(\nu \,t_{1/2})}=-\log_{10}{P}-\log_{10}{P_\alpha} + {\rm constant}\,. 
\end{equation}
In the simplest model, one can start by assuming that both $P_\alpha$ and $\nu$ remain approximately constant, and calculate $-\log_{10}{P}$ based on the premise of a tunneling process. The simplicity of the model brings us to an analytical expression which is then improved by changing the constants to fitted parameters. In this model, the $\alpha$ particle goes through the Coulomb barrier formed due to its interaction with the daughter nucleus. The nuclear potential in this context is treated as a rectangular well with a width of $R_0 = R_\alpha + R_d$, where, $R_\alpha$ and $R_d$ are the radii of the tunneling cluster (in this case an $\alpha$) and the daughter nucleus, respectively. The penetration probability is described by :
\begin{equation}\label{eq:penetration_prob}
    P = \exp\left[-2 \int_{R_{0}}^{R} k(r) \,\rm d r\right]\,,
\end{equation}
where $k\left(r\right)=\sqrt{\frac{2\mu}{\hbar^2}\left|V(r)-E\right|}$ is the wave number within the barrier, and $R$ is the outer classical turning point determined from $V(r)=E$. Here, $E$ corresponds to the energy of the tunneling particle, and $\mu$ is the reduced mass of the emitted $\alpha$-daughter nucleus system. The interaction potential $V(r)$ is composed of contributions from both the Coulomb and centrifugal potentials and is hence given by, 
\begin{equation}
    V(r) = Z_\alpha Z_d\frac{e^2}{r} + \frac{\hbar^2}{2\mu}\frac{\ell(\ell+1)}{r^2}\,.
\end{equation}
 
Replacing for $V(r)$, the wave number is given by $k(r)=\sqrt{\frac{2\mu E}{\hbar^2}\left[\left(\frac{R}{r}-1\right)+\sigma\frac{1}{r^2}\right]}$. Here $\sigma=\frac{\hbar^2}{2\mu E}\ell(\ell+1)$ which represents a small quantity due to the significant difference in scale between the centrifugal and Coulomb potentials \cite{Gamow1949,Zhang-etal2011}. Additionally, in the case where the potential barrier is relatively wide (i.e., $R >> R_0$), we can approximate the integral from Eq.~(\ref{eq:penetration_prob}) as follows
\begin{multline}\label{eq:int_k-l}
    \int_{R_0}^{R}k(r)\,dr=\sqrt{\frac{2\mu E}{\hbar^2}}R\left[\frac{\pi}{2}-2\left(\frac{R_0}{R}\right)^{1/2}\right]\\
    +\sigma\sqrt{\frac{2\mu E}{\hbar^2}}\frac{1}{\sqrt{R_0R}}\,.
\end{multline}

\begin{figure*}[ht]
    \centering
    \includegraphics[scale=0.65]{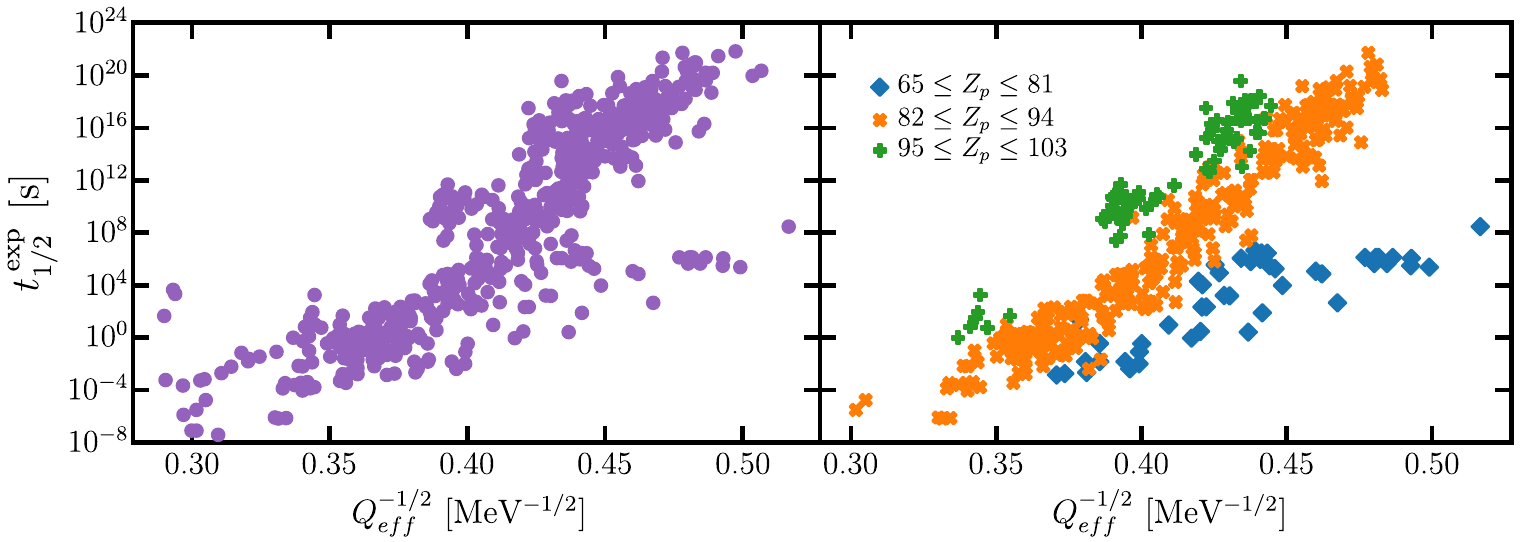}
    \caption{{\em Left panel}: Measured $\alpha$ decay half-lives \cite{NNDC} for 592 decays, where the parent nucleus, daughter nucleus, or both can have multiple excited states associated with the $\alpha$ decay process. The dataset includes parent atomic numbers within the range $65 \le Z_p \le 103$. {\em Right panel}: Dataset divided into three regions of parent atomic numbers and including the condition $\left|\chi'\,\Delta E^{*}/{Q_\alpha}\right|\lesssim15.2$ MeV${}^{-1/2}$.}
    \label{fig:halflife-Qeff1}
\end{figure*}

To extend the $\alpha$ decay half-life calculations to excited states, we begin by defining an effective $Q$ value. Energy conservation allows us to write $m_p + E_p^* = m_{\alpha} + m_d + E_d^* + E_{\alpha} + E_{recoil}$, where $E_p^*$ and $E_d^*$ are the excitation energies of the parent and daughter respectively. With the recoil energy of the heavy daughter being negligible, $E_{\alpha} = Q_{\alpha} + E_p^* - E_d^*$, where, $Q_{\alpha}$ is the usual definition of the $Q$-value of an $\alpha$ decay from a ground-state (g.s.) parent to a daughter in the ground-state. Since the $Q$ value is usually taken to be the tunneling energy in ground state decays, here we define an effective $Q$-value
\begin{equation}\label{eq:Q_eff}
    Q_{eff}\equiv Q_\alpha+E_p^{*}-E_d^{*}=Q_\alpha+\Delta E^{*}\,,
\end{equation}
as the energy of the tunneling $\alpha$. $\Delta E^{*}$ is the difference between the excitation energies of the parent and daughter. The penetration probability, Eq.~(\ref{eq:penetration_prob}), can then be expressed as:
\begin{equation}\label{eq:logP-l-depen_temp}
    \log_{10}{P} = -\left(\beta_1-\beta_2\frac{\Delta E^{*}}{Q_\alpha}\right)\,\chi'-\beta_3\,\rho'-\beta_4\,\frac{\ell(\ell+1)}{\rho'}\,.
\end{equation}
In this equation, the coefficients are defined as $\beta_1=\frac{\pi e^{2}\sqrt{2m_0}}{\hbar\ln 10}$, $\beta_2=\frac{\beta_1}{2}$, $\beta_3=-\frac{4e\sqrt{2m_{0}r_{0}}}{\hbar\ln 10}$, and $\beta_4=-\frac{2\hbar}{\ln10 e\sqrt{2m_0r_0}}$, and the functions $\chi'$ and $\rho'$ are defined as 
\begin{equation}\label{eq:chi}
    \chi'=Z_\alpha Z_d\sqrt{\frac{A_\alpha A_d}{A_pQ_\alpha}}
\end{equation}
and 
\begin{equation}\label{eq:rho}
    \rho'= \sqrt{Z_\alpha Z_d\frac{A_\alpha A_d}{A_p}\left(A_\alpha^{1/3}+A_d^{1/3}\right)}\,,
\end{equation}
respectively.

We observe that the first two terms of the decay law in Eq.~(\ref{eq:logP-l-depen_temp}) are similar to those obtained in Ref. \cite{Qi-etal2009}, specifically when $E_p^{*}=E_d^{*}=0$. These terms are attributed to the Coulomb interaction, representing a fundamental aspect of the decay process. However, the decay law incorporates a third term, which serves to correct for the influence of angular momentum carried by the emitted particle \cite{Qi-etal2012}.

In addition to the above considerations, the impact of unpaired protons and neutrons on the $\alpha$ decay half-life is accounted for in the decay law. This adjustment involves the inclusion of a blocking term, which is not derived from the simple model, but is given as in Ref. \cite{Na_etal2023}. This blocking term, denoted as $\delta_{oe}$, takes the value of 2 for odd-$Z$ odd-$N$ nuclei, 1 for odd-$A$ nuclei, and 0 for even-$Z$ even-$N$ nuclei. Incorporating the outcomes from Eq.~(\ref{eq:logP-l-depen_temp}) and the blocking term, the decay law, which varies with excitation energy, can be expressed as:
\begin{equation}\label{eq:exc_energy_decay_law}
    \log_{10}\,{t_{1/2}^{\rm{\,exc}}}= \beta_1\,\chi'+\beta_2\,\chi'\,\frac{\Delta E^*}{Q_\alpha}+\beta_3\,\rho'+\beta_4\,\frac{\ell(\ell+1)}{\rho'}+\beta_5\,\delta_{oe}\,.
\end{equation}
where the constants $\beta_1$, $\beta_2$, $\beta_3$, and $\beta_4$ have been defined earlier.  Though the constants can in principle be calculated, they are treated as adjustable parameters and fitted to reproduce the $\alpha$ decay half-lives, to compensate for the approximations made during the derivation of the decay law. It is important to emphasize that the objective of the present work is not to fit the universal decay law for ground-state to ground-state decays as is often done in literature but rather to derive an expression suitable for excited nuclei. The contribution of the assault frequency $\nu$ can be incorporated into the other terms of Eq.~(\ref{eq:exc_energy_decay_law}), given its slight variations in the context of $\alpha$ decay compared to the changes in penetration probability.

\section{Fitting the excitation energy-dependent decay law}\label{Fittingdata}
In keeping with the objective mentioned above, we begin by examining in Fig.~\ref{fig:halflife-Qeff1}, the $\alpha$ decay half-lives, $t_{1/2}^{\rm{\,exp}}$, of excited parent nuclei to daughters in the excited or ground state, as a function of $Q_{eff}$ defined in the previous section.

\subsection{Choice of data}
The data in Fig.~\ref{fig:halflife-Qeff1} refer to heavy nuclei with the atomic number $Z_p$ ranging from 65 to 103. At first glance, the data appears quite scattered, making it challenging to perform a reliable fit to derive a decay law (left panel). However, upon closer inspection, we identify a central band of half-lives increasing with $Q_{eff}^{-1/2}$ surrounded by islands. Re-plotting the same data by dividing it into three regions of $Z_p$, as shown in the right panel of Fig.~\ref{fig:halflife-Qeff1}, it becomes clear that the central band corresponds to the region of the $Z_p$ range between 82 and 94. Since most of the considered data falls within this band, we shall focus on this region for performing fits and deriving an excitation energy-dependent decay law. 

\begin{figure}[h!]
    \centering
    \includegraphics[scale=0.6]{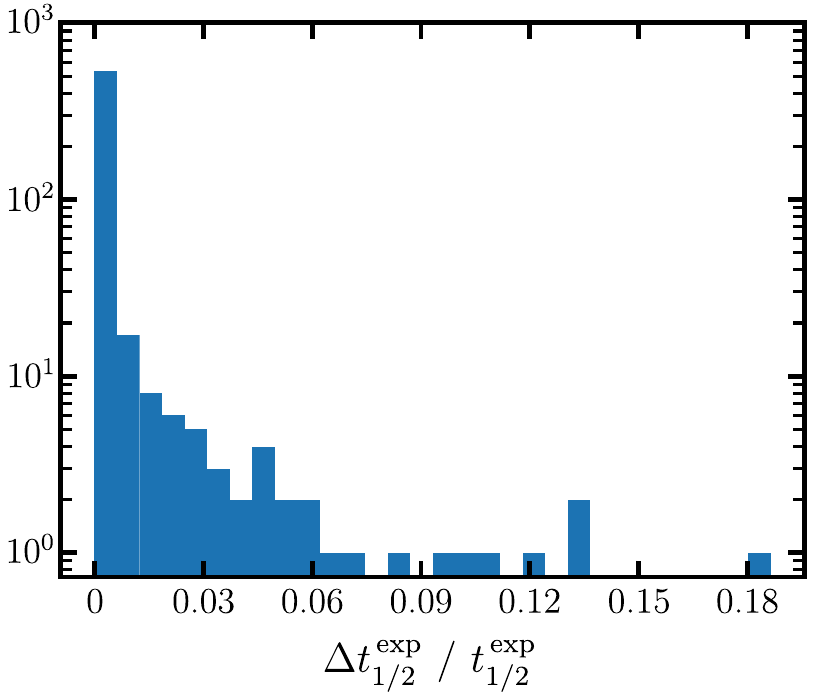}
    \caption{Ratio of uncertainty and the half-life plotted on a logarithmic scale. The vertical axes account for the number of decays with that uncertainty.}
    \label{uncertainties}
\end{figure}

Next, we collected all available data on the uncertainties in the half-lives of the ground and excited states. In Fig. \ref{uncertainties}, we plot the ratio of the uncertainty to the half-life, with histograms indicating the number of decays at each value. In general, the uncertainties are quite small for most of the cases studied. However, it is important to note that in some cases the uncertainties are not listed. To perform a fit that includes these uncertainties in half-lives, we would have to arbitrarily assign error bars in such instances. To test the relevance of the uncertainties, separate fits using the upper and lower limits of the half-lives based on the listed errors were performed. The resulting MSE and $R^2$ values showed negligible changes compared to those obtained using the central values. Consequently, we decided to present the fit results using the central values of the half-lives, as is commonly done in the literature. 

\subsection{Parameters of the decay law}\label{parameters}
Linear regression is a fundamental statistical method employed to analyze the relationships between input and output variables, assuming a linear relationship between a linear combination of input variables ($X_j$) and a single output variable ($Y$). The model for multiple linear regression can be expressed as
\begin{equation}\label{eq:linear_regression}
    Y=\beta_0 + \beta_1 X_1 + \beta_2 X_2 + \cdots + \beta_p X_p
\end{equation}
where $X_j$ represents the $j$th predictor, and $\beta_j$ represents the corresponding model parameter. To establish a connection between the excitation-energy-dependent decay law, Eq.~(\ref{eq:exc_energy_decay_law}) and the linear regression model, Eq.~(\ref{eq:linear_regression}), we consider the logarithm of the half-life $\log_{10}\,{t_{1/2}^{\rm{\,exc}}}$ as the output variable. According to Eq.~(\ref{eq:exc_energy_decay_law}), five predictors are identified, namely $X_1=\chi'$, $X_2=\chi'\,\Delta E^{*}/Q_\alpha$, $X_3=\rho'$, $X_4=\ell(\ell+1)/\rho'$, and $X_5=\delta_{oe}$, together with a constant term $\beta_0$. These depend on basic decay-related information, including the atomic number, mass number, $Q$-value, angular momentum of the $\alpha$ particle, and the excitation energy of the parent and daughter nuclei. This information has been collected from the National Nuclear Data Center (NNDC) \cite{NNDC} for a comprehensive dataset comprising 592 $\alpha$ decays. The nuclei involved in these decays range from $Z=65$ to $103$, with the parent and daughter nuclei existing in either ground-state or excited states. However, we have excluded cases in which the emission occurs from the ground state of the parent to the ground state of the daughter nucleus because this study has been made previously by other authors \cite{GeigerNuttall1911,Royer2000,Soylu2021}. Additionally, to ensure a linear trend in the data, we have divided the dataset into three subsets based on the atomic number of the parent nucleus, namely $65\leq Z_p \leq 81$, $82\leq Z_p \leq 94$, and $95\leq Z_p \leq 103$ (right panel of Fig.~\ref{fig:halflife-Qeff1}). From these subsets, we have chosen to focus on the second set, which comprises 342 cases as explained in the previous section.

\begin{figure}[h!]
    \centering
    \includegraphics[scale=0.55]{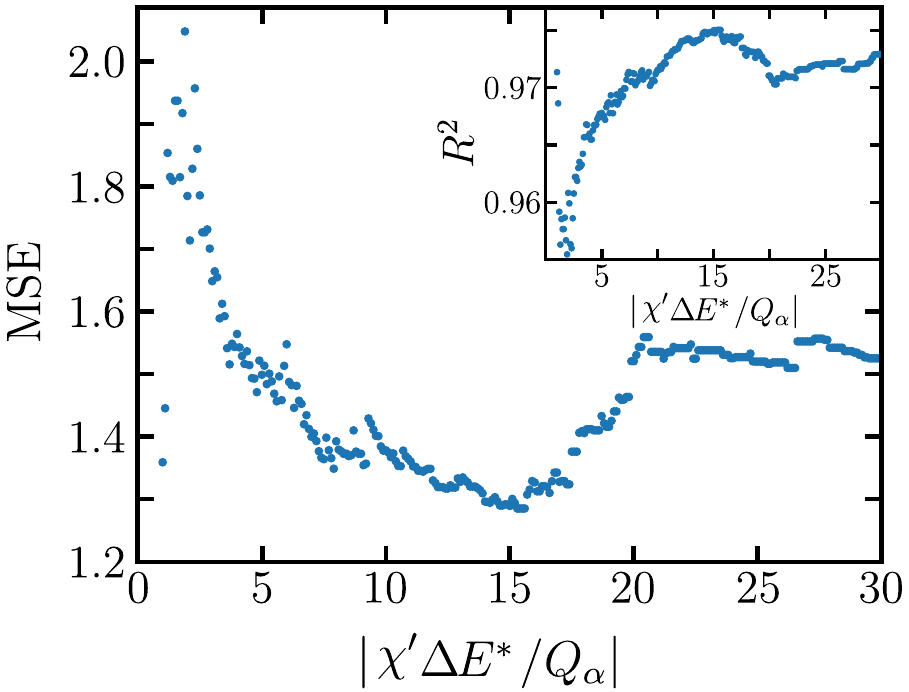}
    \caption{Plot of Mean Squared Error (MSE) versus the quantity $\chi'\,\Delta E^{*}/Q_\alpha$ ranging from 0 to 30 MeV$^{-1/2}$. The inset shows the $R^2$ values vs the same quantity. The MSE reaches a minimum and $R^2$ reaches a maximum at approximately 15.2 MeV$^{-1/2}$, indicating the optimal constraint value for the model.}
    \label{MSE_R2_condition}
\end{figure}

As a final consideration, we examined possible constraints related to the novel term $\chi'\,\Delta E^{*}/Q_\alpha$. To investigate this, linear regression was performed by varying $\left|\chi'\,\Delta E^{*}/Q_\alpha\right|$ from 0 to 30 MeV$^{-1/2}$. Typically, the mean squared error (MSE), which calculates the average of the squares of the differences between the estimated and actual values, is used to identify the best-fitting line representing the relationship between the features and output. As shown in Fig. \ref{MSE_R2_condition}, the MSE reached a minimum at the value of 
$\chi'\,\Delta E^{*}/Q_\alpha$ around 15.2 MeV$^{-1/2}$, while $R^2$ simultaneously reached its maximum value. The fits presented in this work were performed with the 
constraint, $\left|\chi'\frac{\Delta E^{*}}{Q}\right|\lesssim 15.2$ MeV$^{-1/2}$, 
When performing a calculation without any constraints, we found that the quality of the fit reduced slightly. For example, with the constraint, the MSE and $R^2$ values were 1.04 and 0.98, respectively, whereas, without the constraint, they changed to 1.38 and 0.976. Despite this small variation, we decided to continue our analysis with the constraint applied, as it provided a marginally better fit.  

Training our linear regression model from data involves using the Ordinary Least Squares (OLS) technique. OLS minimizes the sum of squared differences between the experimental half-lives and predicted values, providing the best-fit line that quantifies the linear relationship between the predictors $X_j$. To enable a direct comparison of the relative importance of these predictors, we convert all of them to have zero mean and unit standard deviation. This standardized OLS is implemented using the standard Python libraries, specifically {\em scikit-learn} \cite{scikitlearn}.

\begin{table}[h!]
    \caption{Coefficients $\beta_j$ ($j=0,1,2,3,4,5$) and their corresponding standard errors obtained from an OLS fit of Eq.~(\ref{eq:linear_regression}) using experimental data for $\alpha$-decay in nuclei with $82\le Z_p\le 94$. These data points are depicted by orange cross markers in Fig.~\ref{fig:halflife-Qeff1}.}
    \label{tab:coeffs_Linear_Reg}
    \begin{ruledtabular}
    \begin{tabular}{cdd}
        Predictors $X_j$ & \multicolumn{1}{c}{Coefficients $\beta_j$\footnote{The units of $\beta_1$ and $\beta_2$ are in MeV${}^{1/2}$.}} & \multicolumn{1}{c}{Standard error} \\
        \hline
        Constant & 7.1991 & 0.055 \\
        $\chi'$ & 7.4737 & 0.095 \\
        $\chi'\,\Delta E^{*}/Q_\alpha$ & -1.1852 & 0.058 \\
        $\rho'$ & -0.6099 & 0.094 \\
        $\ell(\ell+1)/\rho'$ & 0.6414 & 0.056 \\
        $\delta_{oe}$ & 0.6050 & 0.068
    \end{tabular}
    \end{ruledtabular}
\end{table}

Table~\ref{tab:coeffs_Linear_Reg} displays the results derived from the linear regression, revealing the model parameters $\beta_j$ and their corresponding standard errors. Notably, the analysis underscores that the variables $\chi'$ and $\chi'\Delta E^{*}/Q_\alpha$ are the dominant predictors. This is expected for the former term since this contains the $Q_{eff}^{-1/2}$-dependent behavior predicted by the Geiger-Nuttall law. The second term in Eq.~(\ref{eq:exc_energy_decay_law}) contains information about the excitation energy of either the parent, daughter, or both nuclei. In contrast, $\rho'$, the angular momentum-dependent term, and the blocking term contribute with comparable relative weight to predicting the half-life. The latter also underscores the significance of angular momentum and the importance of considering unpaired protons and neutrons in the decay process.

\subsection{Correlation Analysis}\label{matrixcorrelation}

\begin{figure}[h!]
    \centering
    \includegraphics[scale=0.6]{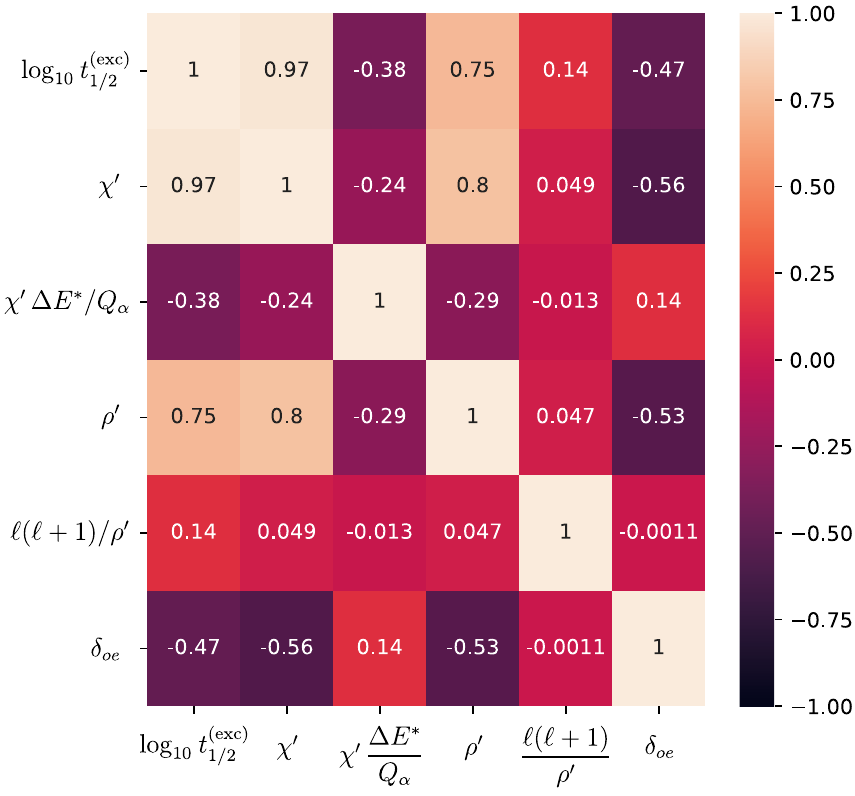}
    \caption{Correlation matrix of target and predictors. Lighter colors represent a positive linear relation between variables, while darker ones indicate an inverse linear relationship. The diagonal line displays perfect correlation, as each variable is perfectly correlated with itself.}
    \label{fig:Correlations}
\end{figure}

Completing the regression analysis, the correlation matrix was generated to obtain a comprehensive overview of the linear relationships between variables in our model. Each entry in Fig.~\ref{fig:Correlations} corresponds to a correlation coefficient, where positive values suggest a positive linear relationship, negative values indicate a negative linear relationship and values closer to zero imply weaker or negligible correlations. Fig.~\ref{fig:Correlations} displays a strong positive correlation between $\log_{10}\,{t_{1/2}^{\rm{\,exc}}}$ and predictors $\chi'$ and $\rho'$ (approximately 0.97 and 0.75, respectively), indicating that changes in these variables are closely aligned with changes in $\log_{10}\,{t_{1/2}^{\rm{\,exc}}}$. Conversely, a significant negative correlation of approximately $-0.47$ and $-0.38$ exists with $\delta_{oe}$ and $\chi'\Delta E^*/Q_\alpha$, respectively, suggesting an inverse linear relation. Additionally, the correlation with $\ell(\ell+1)/\rho'$ indicates a weak linear association between the $\log_{10}\,{t_{1/2}^{\rm{\,exc}}}$ and this predictor.

\begin{figure*}[ht]
    \centering
    \includegraphics[scale=0.55]{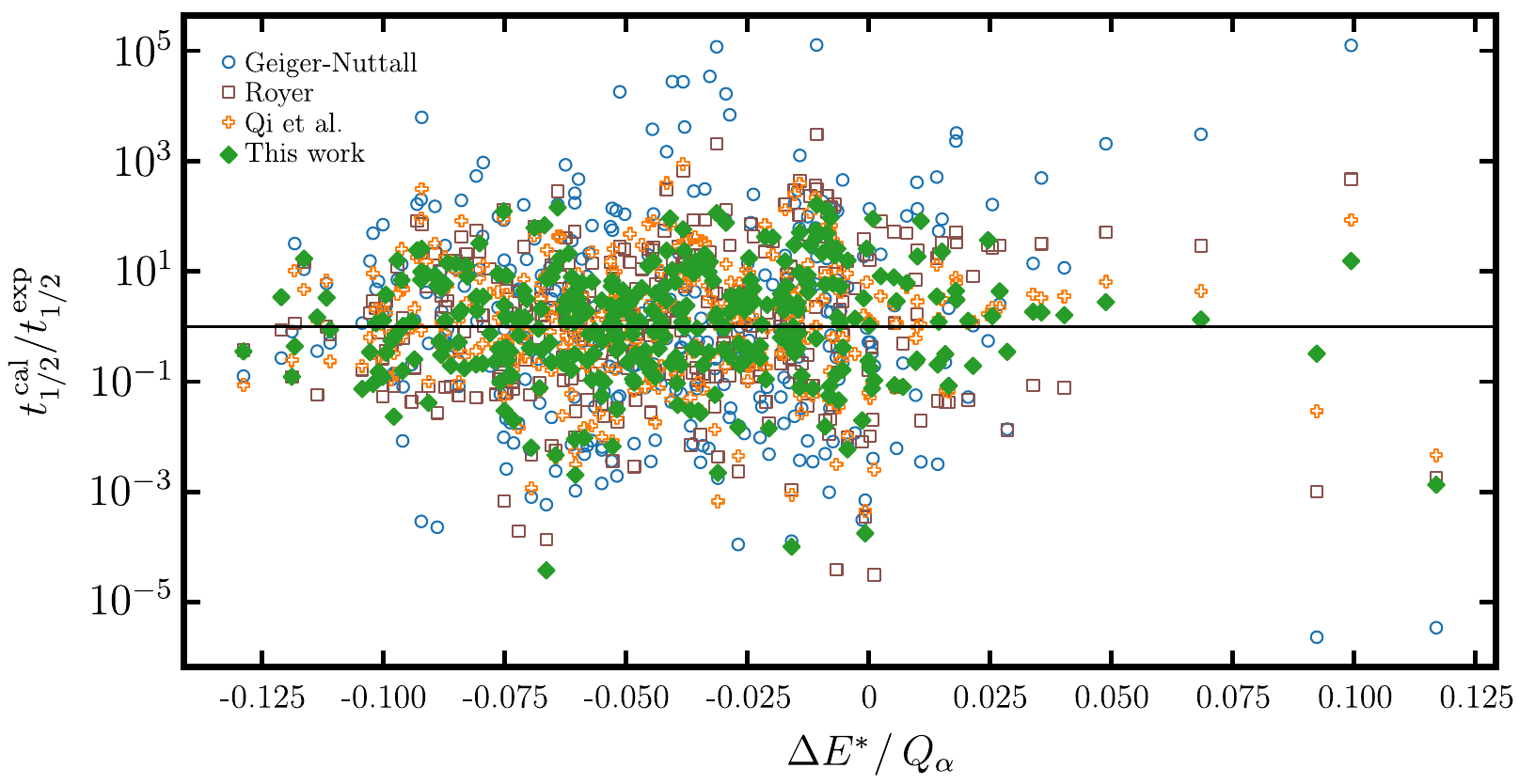}
    \caption{Scatter plot of the ratio of the experimental and calculated empirical half-lives using the fitted Geiger-Nuttall law (circles), re-fitted formulas of Royer (squares) and Qi {\it et al.}, (open diamonds) and the fitted decay law of the present work (filled diamonds) as a function of the factor $\Delta E^*/Q_{\alpha}$ in Eq.~(\ref{eq:exc_energy_decay_law}) which reflects the ratio of the effective and ground state $Q$ values.}
    \label{fig:Scatterplot}
\end{figure*}

To address potential multicollinearity issues in our model, we first examined the correlation coefficients between pairs of predictors. Notably, a strong positive correlation was observed between $\chi'$ and $\rho'$, suggesting potential redundancy in the information provided by these predictors, as expected from their definitions, Eqs.~(\ref{eq:chi}) and (\ref{eq:rho}). To mitigate this redundancy, we applied Ridge and Lasso regression techniques. Remarkably, despite the presence of multicollinearity, neither $\chi'$ nor $\rho'$ were excluded from the model after applying Ridge and Lasso penalties. However, it is interesting to note that the predictor $\rho'$ switched to a negative correlation with half-life, while $\chi'$ remained positive and was the most important feature. Furthermore, evaluation of the $r$-squared and MSE metrics yielded values of 0.976 and 1.08, respectively. These results suggest that while the regularization techniques did not lead to the exclusion of predictors, they also did not significantly improve the model's predictive ability or overall performance. To diminish this redundancy, we also tested the model by dropping the $\rho'$ predictor and computing again the linear regression. In this redefined model, $R$-squared is 0.975, and the MSE is 1.17, indicating that the model remains a good approximation even after addressing the potential multicollinearity issue. For a detailed comparison with other models, please refer to Table~\ref{tab:R2_MSE_models}.

\subsection{Modified decay laws with effective $Q$ values}\label{decay-laws}
The main objective of this work is to fit the data on decays of excited nuclei by an excitation energy-dependent decay law as given in Eq.~(\ref{eq:exc_energy_decay_law}). However, there exist several works in literature that provide empirical formulae that have been fitted using the decays of nuclei in their ground state. Here, we shall consider these formulae from literature and re-fit the parameters in them using an effective $Q$ value as in Eq.~(\ref{eq:Q_eff}) and the data on excited nuclei considered in the present work. 

The different decay laws considered for re-fitting are: 
\begin{enumerate}
    \item[(i)] Geiger-Nuttall \cite{GeigerNuttall1911} law:   
    \begin{equation}\label{eq:GeigerNuttal_law}
        \log_{10} t_{1/2}^{\rm(GN)} = a\,Q_{eff}^{-1/2}+b\,,
    \end{equation}
    with $Q$ in Eq.~(\ref{GeigerN}) replaced by $Q_{eff}$. 
    \item[(ii)] G. Royer \cite{Royer2000} performed fits by assuming that the ground state half-lives depend on the $Q$-value of the decay, the mass of the parent nucleus, and its charge. Here we re-fit the decay law in \cite{Royer2000} by replacing $Q_{\alpha}$ by $Q_{eff}$ as follows:
    \begin{equation}\label{eq:Royer_formula}
        t_{1/2}^{\rm(Royer)} = a\,Z_pQ_{eff}^{-1/2}+b\,A_p^{1/6}Z_p^{1/2}+c\,.
    \end{equation}
    \item[(iii)] The half-lives of nuclei decaying by the emission of a wide range of charged particles has been explained by a universal decay law (UDL) in \cite{Soylu2021}. Once again replacing $Q_{\alpha}$ amounts to changing $\chi^{'}$ in \cite{Soylu2021} to $\chi^{'}_{eff}$, such that 
    \begin{multline}\label{eq:Qietal_formula}
        \log_{10} t_{1/2}^{\rm(Qi)} = a\,\chi^{'}_{eff}+b\,\rho'+\,c+d\,\rho'\sqrt{\ell(\ell+1)}\\+e\,\sqrt{I_p(I_p+1)}+f\,A_p\left[1-(-1)^\ell\right]\,
    \end{multline}
    with 
    \begin{equation*}
        \chi^{'}_{eff} = Z_\alpha Z_d\sqrt{\frac{A_\alpha A_d}{A_p Q_{eff}}}\, .
    \end{equation*}
    The first two terms of Eq.~(\ref{eq:Qietal_formula}) dominate the Coulomb penetration, while the last three ones include the angular momentum and isospin dependence.
\end{enumerate}

\subsection{Comparison of the fitted decay laws}\label{comparison-laws}
The four distinct linear models, namely, the Geiger-Nuttall law, Eq.~(\ref{eq:GeigerNuttal_law}), the formula of Royer, Eq.~(\ref{eq:Royer_formula}), and the UDL derived by Qi and collaborators, Eq.~(\ref{eq:Qietal_formula}), re-fitted with effective $Q$ values to the half-life data of excited nuclei (for the same data set as in the previous sections), and the excitation-energy-dependent decay law of the present work, as given in Eq.~(\ref{eq:exc_energy_decay_law}) are now compared. To evaluate the quality of the regression model, we employ two metrics: $r$-squared and MSE. While all four models yield $r$-squared scores close to one, the excitation-energy-dependent decay law of the present work demonstrates a better fit compared to the others (see Table~\ref{tab:R2_MSE_models}). Fig.~\ref{fig:Scatterplot} displays the spread of the predicted data for each model. In the case of the MSE, we observe that predictions derived from the Geiger-Nuttall law exhibit a larger spread than those of the other models, indicating a higher degree of variance from the experimental half-lives. In contrast, the excitation-energy-dependent decay law of this work once again emerges as the most accurate model for reproducing the experimental data.

\begin{table}[h!]
    \caption{Comparison of the accuracy of the fit of the modified decay laws with the excitation energy-dependent decay law of the present work.}
    \begin{ruledtabular}
    \begin{tabular}{lcc}
        \multicolumn{1}{c}{Model} & \multicolumn{1}{c}{$R^2$} & \multicolumn{1}{c}{MSE} \\
        \hline
        Geiger-Nuttall law & 0.939 & 3.14 \\
        Royer formula & 0.966 & 1.75 \\
        Qi {\it et al.} UDL & 0.976 & 1.23 \\
        This work & 0.980 & 1.04
    \end{tabular}
    \end{ruledtabular}
    \label{tab:R2_MSE_models}
\end{table}

\section{$\alpha$ decay half-lives at elevated temperatures}\label{temperature}
It is natural to expect that at elevated surrounding temperatures, $\alpha$-emitting excited states of nuclei would be populated, thus affecting the total half-lives. Understanding their contribution could be relevant to nucleosynthetic outcomes in environments where the temperatures are high enough to have an impact. In \cite{Jhoan_2023}, the behavior of half-lives of some parent nuclei when their excited states were populated, as a function of temperature is considered. Specifically, in \cite{Jhoan_2023}, those 5 parent nuclei decaying to daughters at the neutron number, $N$ = 126 shell closure, were chosen due to the large number of excited states decaying by $\alpha$ decay. Here, we extend that work to a large set of nuclei and study the prediction power of our model. We start by fixing the temperature and finding the half-lives using known experimental information. We continue by applying our model of Eq.~(\ref{eq:exc_energy_decay_law}) to many $\alpha$ emitters, and finalize by estimating the validity of our approximation by comparing these two approaches.
\begin{figure}[h!]
    \centering
    \includegraphics[scale=0.6]{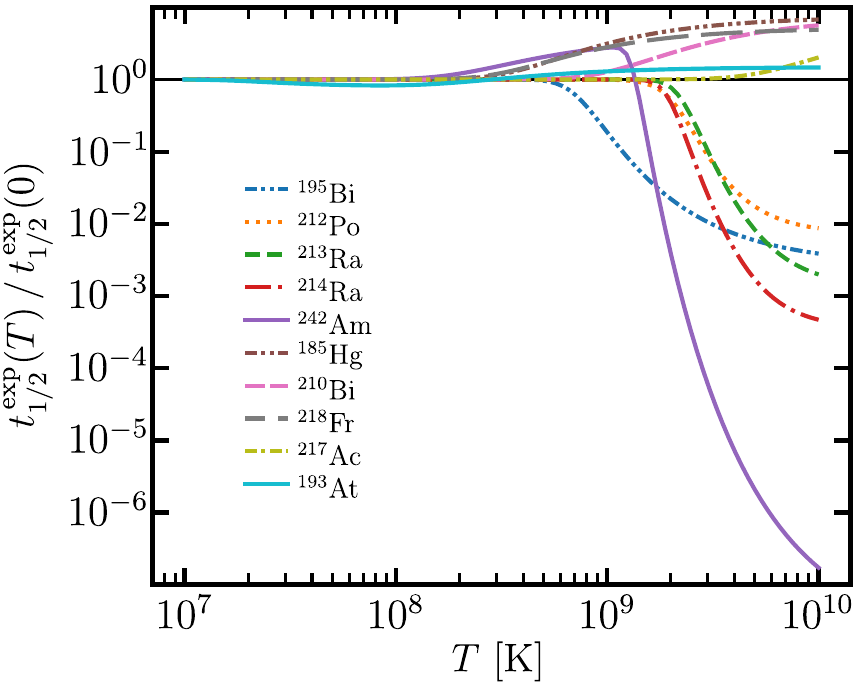}
    \caption{Temperature dependence of the $\alpha$ decay half-lives evaluated as in Eq.~(\ref{eq:half-life-T}) using the available experimental data on $t_{1/2}(E_{p_i}^{*})$ for the $\alpha$ decay of selected nuclei.}
    \label{fig:ratio_halflife_temp}
\end{figure}

\begin{figure*}[ht]
    \centering
    \includegraphics[scale=0.92]{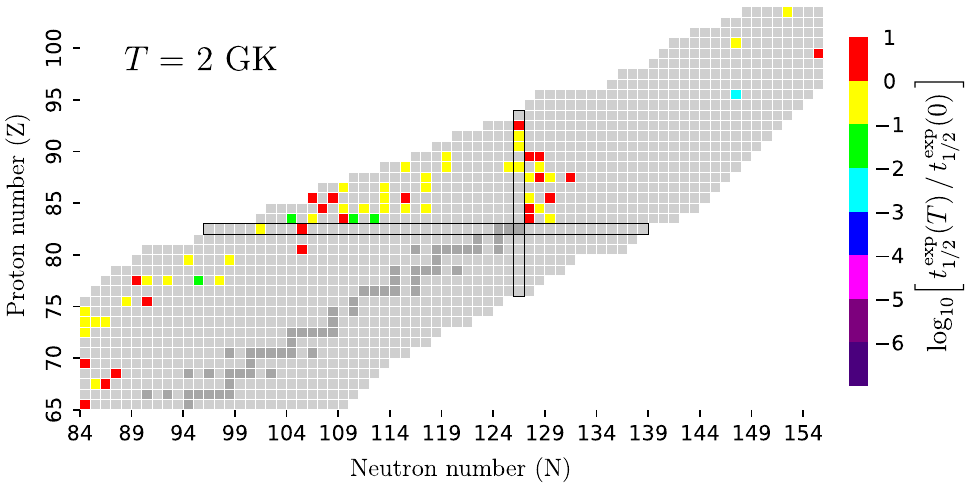} 
    \caption{Temperature-dependent half-lives $t_{1/2}^{exp}(T)$ at $T=2\,\mathrm{GK}$ normalized by those at $T=0$. Only experimental data (with known $\alpha$ branching) has been used as an input in these stellar half-lives given by (\ref{eq:half-life-T}). }
    \label{fig:halflife_T2GK}
\end{figure*}

The stellar $\alpha$ decay half-life is given by \cite{Ward-Fowler1980}
\begin{equation}
    t_{1/2}(T)=\left[\,\frac{1}{\mathcal{G}}\,\sum_{i}\, \frac{g_{p_i}
\exp\left(-E^*_{p_i}/k_BT\right)}{t_{1/2}\left(E_{p_i}^*\right)}\right]^{-1},
\label{eq:half-life-T}
\end{equation}
where $T$ is the ambient temperature, $k_B$ is the Boltzmann constant, and the sum runs over the excited levels of the parent ($p_i$). $\mathcal{G}=\sum_{i} g_{p_i} \exp \left(-E^*_{p_i} / k_B T\right)$ is the partition function, and $g_{p_i}=(2J_{p_i}+1)$ is the statistical weight of the parent's excited state with spin $J_{p_i}$. Above, $t_{1/2}(E_{p_i}^*)$ is the $\alpha$ decay half-life of the parent's excited state $i$. The stellar $\alpha$ decay rate of $^{212}$Po, using the above equation (\ref{eq:half-life-T}) was evaluated and analyzed in great detail regarding each of the experimentally observed excited states decaying by $\alpha$ decay, in \cite{Mohr_2023}. A toy model was also presented to study the role of the contributing factors in (\ref{eq:half-life-T}) to the temperature-dependent $\alpha$ decay half-lives. Using the toy model, the author showed that, as long as the temperature is below $T=2$ GK, it was sufficient to consider the excited levels up to 2 MeV to calculate the temperature-dependent decay rate of $^{212}$Po. In the same spirit and as in \cite{Mohr_2023}, in the next section, we shall investigate the temperature-dependent half-lives of several nuclei using the available data on the $\alpha$ decays of excited states. 

The semi-empirical excitation energy-dependent decay law can be rewritten as \cite{RojasKelkar2022}  
\begin{equation}
\log_{10}[t_{1/2}(E_p^*,E_d^*)] = \log_{10}[t_{1/2}(g.s. \to g.s.)] - \beta_2 
\chi^{\prime} {\Delta E^* \over Q}
\end{equation} 
where $t_{1/2}(g.s. \to g.s.)$ is the half-life when the emission occurs from the ground state parent to the ground state daughter. One can, therefore, consider an enhancement function similar to that in \cite{Perrone1971}, namely, 
\begin{equation}
H = F(E_p^*, T) \times 10^{\beta_2 \chi^{\prime} {\Delta E^* \over Q}} \,.
\end{equation} 
This function will in principle have a similar behavior as in \cite{Perrone1971} since $F(E_p^{*},T)$, which contains the population probability, decreases exponentially and the remaining factor grows very fast leading to a peak in the function $H$ at some excitation energies. However, the situation here is a bit more involved due to the fact that $H$ depends on the excitation energies of the parent and daughter nuclei in contrast to \cite{Perrone1971}, where only decays to the ground state were considered. 

\subsection{Temperature dependence of half-lives from experimental data on excited nuclei}
Taking the available data from \cite{NNDC}, we calculate half-lives as a function of temperature  $t_{1/2}(T)$, for all the experimentally studied $\alpha$ emitters with $65\le Z_p\le 94$. We denote this quantity as $t_{1/2}^{\rm{exp}} (T)$, to indicate that only experimental data has been introduced in the stellar rate (Eq. \ref{eq:half-life-T}). We include levels with known $\alpha$ decay branches only. Fig.~\ref{fig:ratio_halflife_temp} shows the results normalized to the zero temperature values, i.e. ground state half-lives, for some of the nuclei considered. Changes are more noticeable for temperatures beyond 1 GK (note the log scale). While some nuclei present a decrease in their half-lives (as previously found for the 5 nuclei in \cite{Jhoan_2023}), others exhibit an increase.  These enhancements are due to a considerably large spin value of a parent's excited state $J_{p_i}$, and/or a longer half-life of an excited state, both quantities compared to their respective ground state values. Such is the case of, for example, the isomeric state 9$^-$ of $^{210}$Bi, which has a total half-life of $3.06\times 10^{6}$ years, much longer than the 5.012 days half-life of the $1^{-}$ ground state that has a large probability of also decaying by $\beta$ emission. An interesting behavior is that of $^{242}$Am, which has an initial increase in $\alpha$ decay half-life as $T$ departs from zero. That is followed by a continuous decrease for temperatures above $\sim$1.5 GK. At low temperatures, the isomeric 5$^-$ state with a total half-life of 141 years (compared to 16.02 h of the ground state), and a low excitation energy of 0.0486 MeV, can be populated contributing to an increase in half-life. However, higher temperatures populate the state (2$^+$, 3$^-$), at 2.2 MeV, which with a shorter half-life of 13.9 ms leads to the observed decrease.

Despite the interesting features listed above, this analysis is clouded by uncertainties in the data. In the absence of a confirmed $\alpha$ decay, we have used the upper limits provided by the NuDat database \cite{NNDC}. Further uncertainties arise when one compares the NuDat information with the literature. For example, NuDat provides a value of 10$^{-6}$ for the upper limit on the absolute $\alpha$ branch of the $1^-$ ground state but in \cite{aleksandrov} it was found that if there exists an $\alpha$ decay of $^{242}$Am with the $\alpha$ particle energy in the range of 5000 to 5300 keV, its probability is no more than 10$^{-7}$ of the total number of decays of the ground state (g.s.) of $^{242}$Am. Given the range of $\alpha$ particle energies, such decays would correspond to $^{242}$Am (g.s.) decaying to some excited states of $^{238}$Np with excitation energies of about 300 to 500 keV. We chose the NuDat value for the branching fraction and in the absence of information assumed that $^{242}$Am (g.s.) decays to the g.s. of $^{238}$Np. Similarly, there are excited states that can potentially contribute to $\alpha$ emission, but for which there is no experimental information. For example, the $0^-$ state of $^{210}$Bi at energy 46.5 keV could transmute to $^{206}$Tl, however, measuring its branching ratio would be difficult given its electromagnetic half-life of about 3 nanoseconds. If states like this one are indeed $\alpha$ emitters then the half-lives of the isotopes would be reduced. Under these limitations, our work aims at providing estimates of stellar decay rates based only on available experimental data.

A word of caution is in order here. Our results are based on the assumption that thermal equilibrium has been achieved between nuclear states. It has been shown in literature \cite{Ward-Fowler1980} that there exist criteria for limiting temperatures below which the isomeric states for example are not in equilibrium with the ground state and hence may be treated as different species of nuclei \cite{mischApJ}. In the absence of thermal equilibrium, the assumptions regarding the occupation probabilities of nuclear levels applied in this work may be in doubt. While the initial temperature conditions in several astrophysical sites might allow for such an equilibrium, finding an isomer's equilibration temperature, especially when decaying to stability is the most dominant process, requires the knowledge of the specific stellar conditions and the use of a specialized network \cite{mischApJ}. Identifying, astromers, i.e. isomeric states with an impactful role in shaping light curves needs careful consideration \cite{fujimoto}. A detailed study of the $\beta$ decay rates including isomers and the consideration of their thermalization as a function of temperature has been done in literature \cite{mischApJ,mischApJl,Runkle,Reifarth}. Thermal equilibration of high-spin isomers requires transitions via higher-lying intermediate states. This has been discussed in the literature for the examples of $^{180}$Ta and $^{176}$Lu \cite{Hayakawa,MohrPRC79}. 

Details of a method to compute thermally mediated transition rates between the ground states and long-lived isomers has been outlined in \cite{mischApJ,GuptaMeyer}. The authors in \cite{mischApJ}, however, note that most isomers transition to lower energy states preferentially over destruction channels and this connection ensures that destruction cannot cause deviation from thermal equilibrium. Apart from this, isomers which do make a difference, may not have an effect in all environments. An isomer may prevent thermalization at low temperatures but in a hotter environment, thermally driven transitions through intermediate states can enable equilibration. For example, during r-process nucleosynthesis, isomeric states of heavy nuclei can be populated by neutron captures, radioactive decay of a parent nucleus, and photon absorption among other processes. Figure 4 of \cite{mumpower} shows that within a few seconds the heaviest elements, including $\alpha$ emitters, have been synthesized. At that stage, temperatures are high so it is reasonable to assume thermal equilibration between energy levels. After neutron freeze-out, once neutron capture and photodissociation rates are not in equilibrium, radioactive decay shapes the final abundances and determines the evolution of light curves (see  \cite{fujimoto} for the role of isomers in the case of $\beta$ decay). 

Performing such a calculation applying methods as those developed in \cite{mischApJ,mischApJl,GuptaMeyer,fujimoto} is a full-fledged project in itself. In \cite{Sprouse}, while developing a novel decay network Jade that handles nuclear decays and transitions between excited states, the authors note that the complete inclusion of isomers is subject to challenges from a nuclear data perspective. The same is in part true also for the data on $\alpha$ decay. Hence, the half-lives of $\alpha$-decaying isotopes of the present work can serve as a first gauge to determine a rating of isomers that are more influential especially given that not all isotopes with isomeric states will be equally produced (see Ref. \cite{mischApJl} for a rating of $\beta$-decaying isomers). 

In Table \ref{tab:Isomers}, we list the isomeric states of the isotopes that we find to have increased half-lives; some of these count with little or no experimental information. In the table, we highlight those that were also identified in \cite{Misch2024} as a good starting point for experimental searches.

\begin{table}[h!]
    \caption{Isotopes that exhibit longer half-lives with increasing temperature, found in this study. Isotopes marked with ${}^*$ are also listed in Ref \cite{Misch2024}.}
    \label{tab:Isomers}
    \begin{ruledtabular}
    \begin{tabular}{l c l c }
   Isotope &  Isomer $J^\pi$ & total $t_{1/2}^{\rm{\,exp}}$& \%$\alpha$ \\
        \hline
      ${}^{149}$Tb & 11/2${}^{-}$ & 4.17(5) min& 0.02 \\
      ${}^{153}$Ho* & 1/2${}^{+}$ & 9.3(5) min& 0.18 \\
      ${}^{153}$Tm* & 1/2${}^{+}$ & 2.5(2) s& 92 \\
      ${}^{155}$Lu* & 1/2${}^{+}$ & 138(8) ms& 76\\
                    & (25/2${}^{-}$) & 2.69(3) ms& 100 \\
      ${}^{165}$Re* & (11/2)${}^{-}$ &  1.74(6) s& 13\\
      ${}^{166}$Ir & (9${}^{+}$) & 15.1(9) ms& 98.2\\
      ${}^{185}$Hg* & 13/2${}^{+}$ & 21.6(15) s & 0.03\\
      ${}^{187}$Pb* & (13/2${}^{+}$) & 18.3(2) s& 9.5 \\
      ${}^{191}$Po & (13/2${}^{+}$) & 93(3) ms & 96\\
      ${}^{191}$At & (7/2${}^{-}$) & 2.1(+4-3) ms& 100 \\
      ${}^{192}$Bi* & (10${}^{-}$) & 39.6(4) s& 10\\ 
      ${}^{193}$At & (7/2${}^{-}$) & 21(5) ms& 100\\
                   & (13/2${}^{+}$) & 27(+4-3) ms& 24\\
      ${}^{210}$Bi & 9${}^{-}$ & $3.04\times10^{6}(6)$ yr& 100 \\
      ${}^{211}$Po & (25/2${}^{+}$) & 25.2(6) s& 99.98\\
      ${}^{214}$At & 9${}^{-}$ & 760(15) ns& 100\\
      ${}^{216}$Ac & (9${}^{-}$) &  441(7) $\mu$s& 100\\
      ${}^{218}$Fr & (8${}^{-}$, 9${}^{-}$) &  22.0(5) ms& 100\\
      ${}^{217}$Ac & (29/2${}^{+}$) & 740(40) ns& 4.51\\
      ${}^{218}$U & (8${}^{+}$) &  0.56(+26-14) ms& 100\\
      ${}^{242}$Am & 5${}^{-}$ & 141(2) yr& 0.45\\
                   & (2${}^{+}$, 3${}^{-}$) & 14.0(10) ms& $<$ 0.005\\
      ${}^{254}$Es* & 2${}^{+}$ & 39.3(2) h& 0.32 \\
    \end{tabular}
    \end{ruledtabular}
\end{table}

Trends in half-lives can be seen in Fig.~\ref {fig:halflife_T2GK}, where we show the ratio of stellar half-lives to the ground state values across the nuclear landscape. The temperature was fixed at $T=2$ GK. The colored squares correspond to nuclei known to have excited states that decay by $\alpha$'s, the dark-grey squares correspond to stable nuclei, and the light-grey ones are nuclei that do not decay by $\alpha$ emission and/or have no observed excited states emitting $\alpha$'s. At modest temperatures of a few GK, there is a good fraction of the total nuclei that exhibit an increase in half-life by a factor of $\sim$ 10. A larger group of nuclei show decreasing half-lives, with overall changes of an order of magnitude, a few decreasing by two orders of magnitude, and $^{242}$Am with a decrease of 3 orders of magnitude. An asymptotic trend is observed and shows that nuclei with populated isomeric states do not show further increases. Instead of the ground state, isomeric states are the main contributors to the $\alpha$ decay half-life. If such nuclei were to have short-lived $\alpha$-emitting higher energy levels populated, then they would decay faster. However, with the experimental information we have at this point, the isomeric states are responsible for the increase and asymptotic behavior of $\alpha$-decay half-lives at elevated $T$. On the other hand, for nuclei with decreasing half-lives at 2 GK, a trend towards further reduction of their half-lives is expected as increased temperatures would lead to a higher population of the excited $\alpha$ emitting states. 

\subsection{Validating the model for excitation energy-dependent decay law}
\begin{figure}[h!]
    \centering
    \includegraphics[scale=0.59]{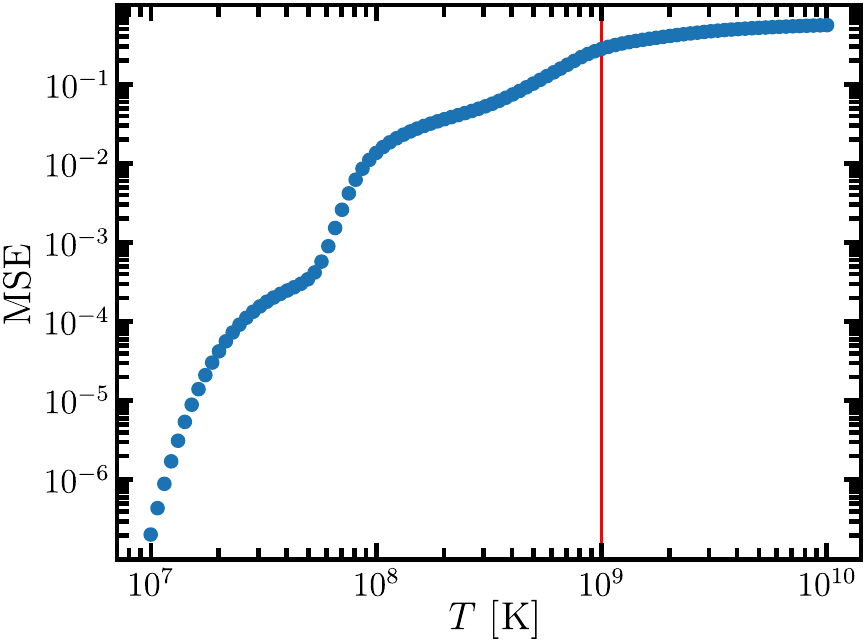}
    \caption{MSE of predicted temperature-dependent half-lives to those based on experimentally known data. All values are normalized to those at $T=0$.}
    \label{fig:MSE_vs_T}
\end{figure}

We now apply the excitation energy-dependent decay law based on a model for $\alpha$ decay of excited nuclei, from the previous sections, to predict the stellar half-lives of thermally excited nuclei. Specifically, we use Eq.~(\ref{eq:exc_energy_decay_law}) to estimate the half-lives of the excited states (with known $\alpha$ decay branches) that enter Eq.~(\ref{eq:half-life-T}) and denote this rate by $t_{1/2}^{\rm{cal}}(T)$. To do so, we first find each nucleus's predicted half-life of excited levels. The calculation included both, decays from the ground-state of the parent to the ground-state of the daughter, and from excited-state to ground- and/or excited-states of the daughter (i.e. we have added to our data in Fig.~\ref{fig:Scatterplot}, decays from parent to daughter ground states). The calculation included 148 decays, in contrast to the fitting done in section \ref{comparison-laws}, which comprised 342 decays. Using our predicted half-lives of excited states, we calculate the half-life temperature dependency, Eq.~(\ref{eq:half-life-T}), for the set of nuclei that satisfy the condition, $\left|\chi'\frac{\Delta E^{*}}{Q}\right|\lesssim 15.2$ MeV$^{-1/2}$. The temperature dependence of the $\alpha$ decay half-lives, $t_{1/2}^{\rm{\,cal}}(T)$, thus calculated is very similar to that in Fig.~\ref{fig:ratio_halflife_temp}. Hence, instead of displaying the calculated $t_{1/2}^{\rm{\,cal}}(T)$ as a function of the temperature, $T$, we rather validate the model by comparing the ratio in Fig.  \ref{fig:ratio_halflife_temp}, namely $R^{\rm{\,exp}}(T) = t_{1/2}^{\rm{\,exp}}(T)/t_{1/2}^{\rm{\,exp}}(0)$ with $R^{\rm{\,cal}}(T) = t_{1/2}^{\rm{\,cal}}(T)/t_{1/2}^{\rm{\,cal}}(0)$ and present the deviation of the calculated numbers, $R^{\rm{\,cal}}(T)$, from $R^{\rm{\,exp}}(T)$ obtained from data, in the form of a plot of the MSE as a function of temperature in Fig.~\ref{fig:MSE_vs_T}. The figure is presented on a log scale to show that the MSE is indeed quite small over a large range of temperatures.
The model for $t_{1/2}^{cal}(T)$ presented here may not necessarily predict the absolute values of temperature-dependent half-lives for all nuclei accurately, however, it can give a good estimate of the change in the half-lives at elevated temperatures for hundreds of nuclei that are involved in the nucleosynthesis path, allowing the community to perform good estimates of related phenomena.   

\section{$\alpha$ decay in different astrophysical scenarios}\label{scenarios} 
Before going over the summary of the findings of the present work, we present the relevance of this work for different types of nucleosyntheses occurring in high-temperature environments.
\begin{itemize}
    \item [1.] One set of nuclei in our study lies in the region of neutron number N $\ge$ 126 and proton number Z $\ge$ 82. Simulations have shown that nuclei in this mass region (see \cite{arcones} and reference therein) and superheavy ones (\cite{petermann, mendoza}, and recently revisited in Holmbeck {\it et al.}(2023) \cite{holmbeck,holmbeck2}) can be synthesized in r-process nucleosynthesis. At the early stages of an r-process, where temperatures start from $\sim$ 10GK and reach 3GK, there is competition between neutron capture,  and photodissociation (and for some conditions with $\beta$ decay), which forms the well-known r-process path below the line of stability.  At later stages, neutrons are exhausted (i.e. at neutron freeze-out) and all the very unstable neutron-rich nuclei, including actinides, and superheavies (if formed) will decay to stability. Simulations identify $\alpha$-decays, especially the decay chains originating from actinide nuclei, as important contributors to the emitted light curve expected from r-process events \cite{2016ApJ...829..110B}.  Depending on the nucleus, there will be competition between decay channels (including fission) and $\alpha$ decay. It is here that the $\alpha$ decay will be of relevance.  
    \item [2.] The above picture, would not be related to the neutron-deficient nuclei that we include in our study (those with 65 $\le$ Z $<$ 82). However, our results could be important for the synthesis of p-nuclei (with 34 $\le$ Z $\le$ 80) via the p-process and the recently proposed $\nu$r-process (see point 3 below).
    
    The abundances of p-nuclei, located on the west side of the stability valley, are still poorly understood \cite{arcones,arnould}. The p-process, (aka $\gamma$-process) which occurs by the dissociation of neutron-rich nuclei via reactions with photons is thought to be responsible for the synthesis of p-nuclei.  The p-process moves matter from nuclei previously produced by the s- and r-process to the proton-rich side of stability \cite{arcones}. It has been discussed that their synthesis has contributions from different astrophysical scenarios, with typical temperatures around 2-4 GK \cite{lambert,choplin}. So far, the path of the p-process is still unclear due to uncertainties in nuclear physics and astrophysical modeling \cite{robert,pignatari}. However, $\alpha$ emission from neutron-deficient nuclei competes with other channels including $\beta$ decay \cite{arcones}.    

    \item[3.] In r-process nucleosynthesis, neutrino interactions also play a key role. During the explosive conditions of an r-process, neutron-rich nuclei can increase their proton number by the capture of neutrinos, in what has been recently proposed as the ``$\nu$r-process" \cite{xiong}. The large neutrino fluxes move matter from the neutron-rich side towards the proton-rich side. Again, unstable nuclei, in this $\nu$r-process path, will decay to stability making $\alpha$ emission at those relevant temperatures.
\end{itemize}

The present study explores $\alpha$ decay in a range of temperatures motivated by the conditions that can be achieved in different astrophysical scenarios. We provide a way to include the temperature dependency of this decay mode. Its application to each of the above nucleosynthesis scenarios is beyond the scope of this study. However, it motivates future work.

\section{Summary and Outlook}\label{conclusion}
Observations of ``kilonovae'' which followed the gravitational waves after neutron star mergers have added fresh impetus to the investigations of different forms of nuclear decays. Neutron-rich matter which emerges from the collision of neutron stars undergoes the r-process nucleosynthesis which is responsible for the production of heavy elements such as gold and platinum. Radioactive decays of the heavy elements power the thermal transients (kilonovae). Models of kilonovae which depend on the energy production from nuclear decays can give us information on the physical conditions during the merger and the aftermath \cite{Metzger2019}. Apart from this, the radioactive decay half-lives are also an essential input \cite{KelkarRojasAPPA} to the networks that predict the abundance of heavy elements.
 
The semi-empirical decay law for half-lives of excited nuclei presented in this work can be used to fill in the missing information on the half-lives of levels that enter such calculations. A good number of the cases that are used in the fits for the excitation energy-dependent decay law consist of the parent in the ground state and the daughter in the excited state. There exist nuclei with a percentage decay to a daughter level, which is around 50 to 100 keV above the ground state, much higher than that to the ground state. For example, $^{224}$Pa has $<$ 0.5\% decay to the ground state of $^{220}$Ac but 70\% decay to an excited state of $^{220}$Ac at 68.7 keV. $^{225}$Th which has 90\% $\alpha$ decay, has only 8.1\% decay to the ground state of $^{221}$Ra but 31\% decay to a level at 321.4 keV. In most cases, the excited daughter decays quickly by emitting a $\gamma$ or series of $\gamma$'s. This situation can change the energy production (distribution between decay modes) scenario of kilonovae, where, if one includes an excitation energy-dependent $\alpha$ decay law, the $\alpha$'s will carry less energy and the photons will carry a bit more. It remains to be tested if this effect would be significant. However, if found important, we note that the photons are also responsible in part for photon-induced fission which in turn could have a small increased contribution.

Given that $\alpha$ decay in the different astrophysical scenarios discussed in the previous section can occur at temperatures of the order of Giga Kelvin, it is crucial to consider the temperature dependence of the nuclear decay rates or half-lives, $t_{1/2}(T)$, of heavy nuclei. The simplest formulation for $t_{1/2}(T)$ involves performing a sum over the half-lives of all excited states of a particular nucleus along with the corresponding population probabilities of the excited states, for a certain kind of decay. The stellar decay rates presented here rely on the simplified assumption of thermal equilibrium between the ground and excited states including isomeric states if any too. The observed features listed below could change if one takes into account the fact that up to a certain value of temperature, isomeric states are not in equilibrium with the ground state of the nucleus. We have investigated the temperature-dependent $\alpha$ decay half-lives, $t_{1/2}^{\rm{\,exp}}(T)$, of heavy elements, using data on 66 cases with the criterion that at least one excited state decays by emitting an $\alpha$. Salient features which arise from the study of $t_{1/2}^{\rm{\,exp}}(T)$ as a function of the temperature, $T$, can be summarized as follows:
\begin{enumerate}
    \item[(i)] Majority of the nuclei studied show a decrease in the half-lives with temperature ($T$), with a good number of cases showing about an order of magnitude decrease at $T$ = 2GK. A few cases with a decrease between 1 and 2 orders of magnitude are also identified.
    \item[(ii)] Exceptional cases where the half-life increases with temperature, however, at the most by an order of magnitude are identified. The increase is associated with the presence of isomeric states.
    \item[(iii)] For $T \gg$ 2 GK, $t_{1/2}^{\rm{\,exp}}(T)$ saturates; however, the decreasing half-lives fall further by an order of magnitude before saturation in contrast to the cases where the increase never exceeds an order of magnitude.
\end{enumerate}

The validity of the model for the decay law of excited nuclei presented earlier is tested by evaluating $t_{1/2}(T)$ using the empirical formula instead of the available data and comparing it with $t_{1/2}^{\rm{\,exp}}(T)$. With an MSE that remains quite small over a large range of temperatures, the empirical decay law presented here can be used in the absence of data. Though out of the scope of the present work, the decay law provided here can be incorporated in nucleosynthesis network codes to study the effects of thermally enhanced (or reduced) $\alpha$ decay rates on the abundance of heavy elements. In passing, we note that the decay law could prove useful for investigations in the field of heavy-ion collisions that involve hot nuclei.

\begin{acknowledgments}
N.G.K. thanks the Faculty of Science, Universidad de Los Andes, Colombia, for financial support through Grant No. INV-2023-162-2841. O.L.C. acknowledges support from the National Sciences and Engineering Research Council of Canada (NSERC).
\end{acknowledgments}


\begin{thebibliography}{39}
    \bibitem{Bethe1936}Bethe, H. \& Bacher, R. Nuclear Physics A. Stationary States of Nuclei. \href{https://link.aps.org/doi/10.1103/RevModPhys.8.82}{{\em Rev. Mod. Phys.}. \textbf{8}, 82-229 (1936)}.
    \bibitem{Weizsacker1935}Weizsäcker, C. Zur Theorie der Kernmassen. \href{https://doi.org/10.1007/BF01337700}{{\em Z. Phys.}. \textbf{96}, 431-458 (1935)}.
    \bibitem{Seeger1961}Seeger, P. Semiempirical atomic mass law. \href{https://doi.org/10.1016/0029-5582(61)90147-X}{{\em Nucl. Phys.}. \textbf{25} pp. 1-IN1 (1961)}.
    \bibitem{Moller1981}Möller, P. \& Nix, J. Nuclear mass formula with a Yukawa-plus-exponential macroscopic model and a folded-Yukawa single-particle potential. \href{https://www.sciencedirect.com/science/article/pii/0375947481904735}{{\em Nucl. Phys. A}. \textbf{361}, 117-146 (1981)}.
    \bibitem{Moller1988}Möller, P. \& Nix, J. Nuclear masses from a unified macroscopic-microscopic model. \href{https://www.sciencedirect.com/science/article/pii/0092640X88900228}{{\em At. Data Nucl. Data Tables}. \textbf{39}, 213-223 (1988)}.
    \bibitem{Moller2016}Möller, P., Sierk, A., Ichikawa, T. \& Sagawa, H. Nuclear ground-state masses and deformations: FRDM (2012). \href{https://doi.org/10.1016/j.adt.2015.10.002}{{\em At. Data Nucl. Data Tables}. \textbf{109-110} pp. 1 - 204 (2016)}.
    \bibitem{Davidson1994}Davidson, N. {\it et al.} A semi-empirical determination of the properties of nuclear matter. \href{https://www.sciencedirect.com/science/article/pii/0375947494902690}{{\em Nucl. Phys. A}. \textbf{570}, 61-68 (1994)}.
    \bibitem{GeigerNuttall1911}Geiger, H. \& Nuttall, J. LVII. The ranges of the $\ensuremath{\alpha}$ particles from various radioactive substances and a relation between range and period of transformation. \href{https://doi.org/10.1080/14786441008637156}{{\em London, Edinburgh Dublin Philos. Mag. J. Sci.}. \textbf{22}, 613-621 (1911)}.
    \bibitem{QiPLB2014}Qi, C. {\it et al.} On the validity of the Geiger–Nuttall alpha-decay law and its microscopic basis. \href{https://www.sciencedirect.com/science/article/pii/S0370269314003761}{{\em Phys. Lett. B}. \textbf{734} pp. 203-206 (2014)}.
    \bibitem{Ren2012}Ren, Y. \& Ren, Z. New Geiger-Nuttall law for $\ensuremath{\alpha}$ decay of heavy nuclei. \href{https://link.aps.org/doi/10.1103/PhysRevC.85.044608}{{\em Phys. Rev. C}. \textbf{85}, 044608 (2012)}.    
    \bibitem{Viola1966}Viola, V. \& Seaborg, G. Nuclear systematics of the heavy elements-II Lifetimes for alpha, beta, and spontaneous fission decay. \href{https://www.sciencedirect.com/science/article/pii/0022190266804128}{{\em J. Inorg. Nucl. Chem.}. \textbf{28}, 741-761 (1966)}.
    \bibitem{Royer2000}Royer, G. Alpha emission and spontaneous fission through quasi-molecular shapes. \href{https://doi.org/10.1088/0954-3899/26/8/305}{{\em J. Phys. G, Nucl. Part. Phys.}. \textbf{26}, 1149-1170 (2000)}.
    \bibitem{Dong2005}Dong, T. \& Ren, Z. New calculations of $\ensuremath{\alpha}$-decay half-lives by the Viola-Seaborg formula. \href{https://doi.org/10.1140/epja/i2005-10142-y}{{\em Eur. Phys. J. A}. \textbf{26}, 69-72 (2005)}. 
    \bibitem{Qi-etal2009}Qi, C., Xu, F., Liotta, R. \& Wyss, R. Universal Decay Law in Charged-Particle Emission and Exotic Cluster Radioactivity. \href{https://link.aps.org/doi/10.1103/PhysRevLett.103.072501}{{\em Phys. Rev. Lett.}. \textbf{103}, 072501 (2009)}.
    \bibitem{Delion2009}Delion, D. Universal decay rule for reduced widths. \href{https://Stellar_alpha_decay_rateslink.aps.org/doi/10.1103/PhysRevC.80.024310}{{\em Phys. Rev. C}. \textbf{80}, 024310 (2009)}.
    \bibitem{Soylu2021}Soylu, A. \& Qi, C. Extended universal decay law formula for the $\ensuremath{\alpha}$ and cluster decays. \href{https://www.sciencedirect.com/science/article/pii/S0375947421000865}{{\em Nuc. Phys. A}. \textbf{1013} pp. 122221 (2021)}.
    \bibitem{Delion2015}Delion, D. \& Dumitrescu, A. Systematics of $\ensuremath{\alpha}$-decay transitions to excited states. \href{https://link.aps.org/doi/10.1103/PhysRevC.92.021303}{{\em Phys. Rev. C}. \textbf{92}, 021303 (2015)}.
    \bibitem{RojasKelkar2022}Rojas-Gamboa, D.~F., Kelkar, N. \& Caballero, O. Temperature dependence of cluster decay. \href{https://doi.org/10.1016/j.nuclphysa.2022.122524}{{\em Nucl. Phys. A}. \textbf{1028} pp. 122524 (2022)}.
    \bibitem{DiegoThesis} Rojas-Gamboa, D.~F.  (2022). Ground- and excited-state calculations of cluster radioactivity and alpha decay. \href{http://hdl.handle.net/1992/64152}{Ph.D. thesis, Universidad de los Andes.}
    \bibitem{Perrone1971}Perrone, F. \& Clayton, D. Thermally enhanced $\ensuremath{\alpha}$-decay and the $s$-process. \href{https://doi.org/10.1007/BF00649638}{{\em Astrophys. \& Space Sci.}. \textbf{11}, 451-462 (1971)}.
    \bibitem{Jhoan_2023}Perez Velasquez, J., Caballero, O. \& Kelkar, N. Alpha decay of thermally excited nuclei. \href{https://dx.doi.org/10.1088/1361-6471/aca03c}{{\em J. Phys. G Nucl. Part. Phys.} \textbf{50}, 015203 (2022)}.
    \bibitem{Mohr_2023}Mohr, P. $\ensuremath{\alpha}$-decay half-life of ${}^{212}$Po at stellar temperatures. \href{https://dx.doi.org/10.1088/1361-6471/acd39a}{{\em J. Phys. G: Nucl. Part. Phys.}. \textbf{50}, 075103 (2023)}.
    \bibitem{Jhoan_2023_2}Perez Velasquez, J., Kelkar, N. \& Caballero, O. Reply to comment on ‘Alpha decay of thermally excited nuclei’. \href{https://dx.doi.org/10.1088/1361-6471/ace63b}{{\em J. Phys. G: Nucl. Part. Phys.}. \textbf{50}, 098001 (2023)}.    
    \bibitem{Gamow1949}Gamow, G. \& Critchfield, C. Theory of atomic nucleus and nuclear energy sources. (Oxford, Clarendon Press.,1949).
    \bibitem{Zhang-etal2011}Zhang, X., Xu, C. \& Ren, Z. $\ensuremath{\alpha}$ decay to members of favored bands in both even-even and odd-A nuclei. \href{https://link.aps.org/doi/10.1103/PhysRevC.84.044312}{{\em Phys. Rev. C}. \textbf{84}, 044312 (2011)}.
    \bibitem{Qi-etal2012}Qi, C., Delion, D., Liotta, R. \& Wyss, R. Effects of formation properties in one-proton radioactivity. \href{https://link.aps.org/doi/10.1103/PhysRevC.85.011303}{{\em Phys. Rev. C}. \textbf{85}, 011303 (2012)}.
    \bibitem{Na_etal2023}Ma, N., Zhao, T., Wang, W. \& Zhang, H. Simple deep-learning approach for $\ensuremath{\alpha}$-decay half-life studies. \href{https://link.aps.org/doi/10.1103/PhysRevC.107.014310}{{\em Phys. Rev. C}. \textbf{107}, 014310 (2023)}.
    \bibitem{NNDC}National Nuclear Data Center (NNDC) Chart of Nuclides. Available at \href{http://www.nndc.bnl.gov}{http://www.nndc.bnl.gov}.
    \bibitem{scikitlearn}Pedregosa, F. {\it et al.} Scikit-learn: Machine Learning in {P}ython. \href{http://jmlr.org/papers/v12/pedregosa11a.html}{{\em JMLR}. \textbf{12} pp. 2825--2830 (2011)}.
    \bibitem{Ward-Fowler1980}Ward, R. \& Fowler, W. Thermalization of long-lived nuclear isomeric states under stellar conditions. \href{https://ui.adsabs.harvard.edu/abs/1980ApJ...238..266W}{{\em Astrophys. J.} \textbf{238} pp. 266--286 (1980)}.
    \bibitem{aleksandrov}Aleksandrov, B.~M. {\it et al.} The decay of ${}^{242}$Am. \href{https://doi.org/10.1007/BF01816912}{{\em At. Energy} {\bf 27}, 724--728 (1969).}
    \bibitem{mischApJ}Misch, G.~W. {\it et al.}, Astromers: {N}uclear {I}somers in {A}strophysics. \href{https://doi.org/10.3847/1538-4365/abc41d}{{\em ApJS} {\bf 252}, 2 (2021).}
    \bibitem{fujimoto}Fujimoto, S. \& Hashimoto, M. The impact of isomers on a kilonova associated with neutron star mergers.    \href{https://doi.org/10.1093/mnrasl/slaa016}{{\em MNRAS} {\bf 493}, L103--L107 (2020).}
    \bibitem{mischApJl} Misch, G.~W., Sprouse, T.~M., \& Mumpower, M.~R. Astromers in the Radioactive Decay of r-process Nuclei. \href{https://doi.org/10.3847/2041-8213/abfb74}{{\em ApJL} {\bf 913}, L2 (2021).}
    \bibitem{Runkle} Runkle, R.~C., Champagne, A.~E., \& Engel, J. Thermal Equilibration of ${}^{26}$Al. \href{https://dx.doi.org/10.1086/321594}{{\em ApJ} {\bf 556}, 970 (2001).}
    \bibitem{Reifarth}Reifarth, R. {\it et al.} Treatment of isomers in nucleosynthesis codes. \href{https://doi.org/10.1142/S0217751X1843011X}{{\em IJMPA} A {\bf 33}, 1843011 (2018).}
    \bibitem{Hayakawa}Hayakawa, T. {\it et al.} Reanalysis of the ($J=5$) state at $592$ keV in $^{180}\mathrm{Ta}$ and its role in the $\ensuremath{\nu}$-process nucleosynthesis of $^{180}\mathrm{Ta}$ in supernovae. \href{https://link.aps.org/doi/10.1103/PhysRevC.82.058801}{{\em Phys. Rev. C} {\bf 82}, 058801 (2010).}
    \bibitem{MohrPRC79}Mohr, P. {\it et al.} Properties of the ${5}^{\ensuremath{-}}$ state at 839 keV in $^{176}\mathrm{Lu}$ and the $s$-process branching at $A=176$. \href{https://link.aps.org/doi/10.1103/PhysRevC.79.045804}{{\em Phys. Rev. C} {\bf 79}, 045804 (2009).}    
    \bibitem{GuptaMeyer}Gupta, S.~S. \& Meyer, B.~S. Internal equilibration of a nucleus with metastable states: ${}^{26}\mathrm{Al}$ as an example. \href{https://link.aps.org/doi/10.1103/PhysRevC.64.025805}{{\em Phys. Rev. C} {\bf 64}, 025805 (2001).}
    \bibitem{mumpower}Mumpower, M.~R. {\it et al.} $\ensuremath{\beta}$-delayed Fission in r-process Nucleosynthesis. \href{https://doi.org/10.3847/1538-4357/aaeaca}{{\em ApJ} {\bf 869}, 14 (2018).} 
    \bibitem{Sprouse}Sprouse, T.~M., Misch, G.~W., \& Mumpower, M.~R. Isochronic Evolution and the Radioactive Decay of r-process Nuclei. \href{https://doi.org/10.3847/1538-4357/ac470f}{{\em ApJ} {\bf 929}, 22 (2022).}
    \bibitem{Misch2024} Misch, G.~W. \& Mumpower, M.~R. Astromers: status and prospects. \href{https://doi.org/10.1140/epjs/s11734-024-01136-z}{{\em Eur. Phys. J. Spec. Top.} {\bf 233}, 1075 (2024).}
    \bibitem{arcones}Arcones, A. \& Thielemann, F.~K. Origin of the elements. \href{https://doi.org/10.1007/s00159-022-00146-x}{{\em Astron. Astrophys. Rev.} {\bf 31}, 1 (2023).} 
    \bibitem{petermann}Petermann, I. {\it et al.} Have superheavy elements been produced in nature?. \href{https://doi.org/10.1140/epja/i2012-12122-6}{{\em Eur. Phys. J. A} {\bf 48}, 122 (2012).} 
    \bibitem{mendoza} Mendoza-Temis, J.~J. {\it et al.} Nuclear robustness of the $r$ process in neutron-star mergers. \href{https://link.aps.org/doi/10.1103/PhysRevC.92.055805}{{\em Phys. Rev. C} {\bf 92}, 055805 (2015).}
    \bibitem{holmbeck} Holmbeck, E.~M. {\it et al.} Superheavy Elements in Kilonovae. \href{https://doi.org/10.3847/2041-8213/acd9cb}{{\em ApJL} {\bf 951}, L13 (2023).}
    \bibitem{holmbeck2}Holmbeck, E.~M., Sprouse, T.~M., \& Mumpower, M.~R. Nucleosynthesis and observation of the heaviest elements. \href{https://doi.org/10.1140/epja/s10050-023-00927-7}{{\em Eur. Phys. J. A} {\bf 59}, 28 (2023).}

    \bibitem{2016ApJ...829..110B} Barnes, J., Kasen, D., Wu, M.-R., et al.\ 2016, \apj, 829, 110. doi:10.3847/0004-637X/829/2/110
    
    \bibitem{arnould}Arnould, M. \& Goriely, S. The p-process of stellar nucleosynthesis: astrophysics and nuclear physics status. \href{https://doi.org/10.1016/S0370-1573(03)00242-4}{{\em Phys. Rep.} {\bf 384}, 1 (2003).}
    \bibitem{lambert}Lambert, D.~L. The p-nuclei: abundances and origins. \href{https://doi.org/10.1007/BF00872527}{{\em The Astron. Astrophys. Rev.} {\bf 3}, 201--256 (1992).}
    \bibitem{choplin}Choplin, A. {\it et al}. The p-process in exploding rotating massive stars. \href{https://doi.org/10.1051/0004-6361/202243331}{{\em A{\&}A} {\bf 661}, A86 (2022).}
    \bibitem{robert} Roberti, L. {\it et al.} The $\ensuremath{\gamma}$-process nucleosynthesis in core-collapse supernovae I. A novel analysis of $\ensuremath{\gamma}$-process yields in massive stars. \href{https://doi.org/10.1051/0004-6361/202346556}{{\em A{\&}A} {\bf 677}, A22 (2023).}
    \bibitem{pignatari}Pignatari, M., G{\"o}bel, K., Reifarth, R., \& Travaglio, C. The production of proton-rich isotopes beyond iron: The $\ensuremath{\gamma}$-process in stars. \href{https://doi.org/10.1142/S0218301316300034}{{\em Int. J. Mod. Phys. E} {\bf 25}, 1630003-232 (2016).}
    \bibitem{xiong}Xiong, Z. {\it et al.} Production of $p$ Nuclei from $r$-Process Seeds: The $\ensuremath{\nu}r$ Process. \href{https://link.aps.org/doi/10.1103/PhysRevLett.132.192701}{{\em Phys. Rev. Lett.} {\bf 132}, 192701 (2024).}
    \bibitem{Metzger2019}Metzger, B. Kilonovae. \href{https://doi.org/10.1007/s41114-019-0024-0}{{\em Living Rev. Relativ.}. \textbf{23}, 1 (2019)}.
    \bibitem{KelkarRojasAPPA}Kelkar, N., Rojas-Gamboa, D.~F., Caballero, O. \& Perez Velasquez, J. Alpha and Cluster Decay in $r$-Process Nucleosynthesis. \href{https://doi.org/10.12693/APhysPolA.142.324}{{\em Acta Phys. Pol. A}. \textbf{142}, 324 (2022)}.
\end{thebibliography}
\end{document}